# Integrated Development Environment Gesture

# for modeling workflow diagrams

Carlos Alberto Fernandez-y-Fernandez[1] and José Angel Quintanar Morales[2]
[1]*Instituto de Computación, LIDIS*[2]
*Universidad Tecnológica de la Mixteca,*
*Huajuapan de León, Oaxaca,* **México**
*{caff,joseangel}@mixteco.utm.mx*

**Abstract**

*The current software development tools show the same form of interaction as when they started back, in the mid 70's. However, since the appearance of visual languages and due to their own nature, they can be handled by tools which have different input methods to conventional ones.*
*By incorporating new motion detection technology, it is intended that new forms of interaction are established. Interactions which respond to the free movement of hands, therefore the software´s developer will have a substantial improvement in the user experience.*

## 1. Introduction

Quality assurance is an activity embedded in the entire process of software engineering, covering aspects such as: methods and tools of analysis, design, coding, testing and formal technical reviews, control of software documentation and changes made, and a monitoring procedure to ensure the software development standards, and mechanisms of measure and information [1].

However, obtaining quality software has greater implications than the simple use of methodologies and procedures and standards for analysis, design, programming and testing software. It is not enough just to standardize the philosophy of work in order to
obtain greater reliability, lower maintenance cost and facilities at the testing stage [2].

Not all aspects of software development have been covered; one of the stages which until now still lacks attention is implementation, an aspect which is fundamental to any methodology or software development process. This is where the software developer builds an approximate solution to a specific problem.

As this is an entirely intellectual activity, special attention should be given to the supporting tools[3], the integrated development environment (IDE) being this one of the most important ones. Despite this, for this tool there is not a significant advance in terms of adaptation and the use of new technologies geared towards improving the user experience [4]. To raise the proposed project is essential to first analyze the origins and trends of the IDE's.

## 1.1 A Background of Integrated Development Environment.

The use of IDE's dates back to the early 70's, it became popular when the need to develop complex software emerged. The construction of the software required a console or terminal and the use of a great number of specialized commands, particularly because the software prior to that time (until the late 60's) was designed using flowcharts and was later implemented through punch cards or paper tape. Therefore, there was a lack of support to be reviewed by a compiler.

It was not until the year 1975 that the first IDE in history, the "Master I" [5] developed by SoftLab Munich was announced.

Master I was in a hybrid arrangement of hardware and software. As a mean of entry it used a keyboard similar to that of today's computers, this way the input of data in a totally intuitive manner was possible. Also, it was possible for the first time to compile (in a very primitive manner) the source code programs. To a great extend this was due the storage code on a disk and not on magnetic tape.

However the real impact was generated by the way developers interact with Master I, It should be remembered that until then programs had be coded using punch cards, followed by a tedious process of inserting codes. These actions were performed manually, introducing capture errors.

This is how market introduction of Master I was considered a milestone mainly because it created a new technology proposal, which took as basis the reduction of complexity to the user through a system of interaction without predecent [5].

Since that time a long evolution to improve the IDE's. started. However, the majority of projects have been focused on a single path. The emergence of new and better technologies [4].

IDE's developers, both private companies and independent projects, appear to have stalled on a single line of development, based on the integration of complex and powerful tools available in modern IDEs.

### 1.1 Features Integrated Development Environment.

Currently an IDE is a software which provides greater capabilities for developers to implement software, an IDE integrated traditionally:

- A text editor for source code.
- Integration of a compiler and / or interpreter, or with the ability to communicate with a compiler.
- Tools for automated building code, such as auto and prediction of reserved words.
- Tools for debugging code.

Sometimes a version control system is integrated as well as other tools that simplify the construction of user interfaces.

Today most modern IDE's incorporate powerful classes viewers, inspectors of objects, inheritance hierarchy diagrams, particularly for developing object-oriented systems [1].

The incorporation of the various tools mentioned above, continue to focus on the same line of evolution, on the main points of automation and simplification of writing code. This does not mean that the efforts made by different development teams IDE's are wrong, or are not substantiated, since its objective is to build more robust IDE's.

Only recently have the projects have taken alternative branches, as in the case of Eclipse, which departs from the traditional scheme of specialized IDE, and which is defined as:

> "a kind of universal tool - an open extensible IDE for anything and everything in particular"[1]

Even with the great advances, especially in the architecture of Eclipse and portability features, it does not manage to get out of the previously established patterns of interaction.

In the late nineties the first efforts to develop visual languages appeared, Although there was support for most of them, their own abstract nature, offers the possibility to change the traditional way of interaction [6].

### 1.2 Visual languages.

With the development and standardization of the Unified Modeling Language (UML) the first steps towards visual programming were taken, something similar to the way software used to be developed in the late 60's [7], but now with more technology. With the release of more robust hardware it was possible to automatically translate and compile visual models directly.

Due to the increased use of visual languages, the first IDE's with visual models translators appeared. These IDE's have the total capacity of generating code from the visual representation of a model, such as translating a UML class diagram to its corresponding code in a specific language (either Java or C++) [8].

There are environments with support for specific visual languages. The most remarkable we have is the Robotic

---

[1] The Eclipse Foundation: http://eclipse.org/

Investion System (RIS) marketed as Lego Mindstorm. Basically it's an environment that allows a microcontroller to handle different sensors. The operation is carried out by the arrangement of icons, which are translated and subsequently downloaded to the NXT (ARM7 microcontroller 32-bit).

The importance of RIS lies in the ease of use, to such an extent that its segment of users ranges from children aged 6 years to adult users, who are interested in starting in the area of software development [9].

A similar project is Scratch, unlike Lego Mindstorms, Scratch projects are built with objects (in this case is the translation that the team gives to sprite) based on the instructions that are programmed for the object, they are composed of brightly colored graphical blocks, batteries form which gives them the script name. Its interface and the concept of blocks to implement the code make it very attractive and understandable for children and teenagers [10].

Another IDE for visual languages, which is also the one with the most consolidation is the Virtual Instrumentation Engineering Workbench Laboratory (known commercially as LabVIEW). It is a development platform for a visual environment developed by National Instruments, this IDE provides the necessary tools to obtain a layer of data where the signals and related data allowed by libraries are acquired. Usually the development of such specialized applications, involves the use of complicated languages on complex architectures, such as VHDL or assembler. The use of LabView minimizes development time and complexity of the applications themselves. LabView has been used in building the particle accelerator at CERN (European Organization for Nuclear Research) [11].

Despite the existence of different visual languages, UML is the one which dominates the market, as a similar alternative method we have the Discovery. Discovery Method is a methodology for object-oriented development formally proposed in 1998 by Anthony J H. Simons [12]. Its diagram of tasks pertaining to the business modeling phase is an entity visually similar to an activity diagram of UML, with the advantage of having a formal semantic representation.

In contrast to UML, Discovery has a limited number of CASE tools, as such, is insufficiently diffusion among software developers. There is currently a plug-in for Eclipse, which facilitates the creating of a task diagrams and its respective translation task to their respective task algebra.

## 1.3 Local applications and online applications.

Master I like the vast majority of current IDEs, runs locally, ie, you have to download and then install it, this requires that the software has support for the specific platform.

The tendency to build online applications and their respective use by users has had a positive impact [13]. However the increase in data transmission speed any IDE is adapted to behavior on the web. One tool that shows a Picnik site is best behavior.

Picnik is a kind of website editor online image graphics, with the ability to crop, resize, and insert special effects as well as having other advantages such as patterns and styles of graphic designers, all in real time and with superior portability to the current platform applications. Although it is not an IDE as such, it is an example of how it could become an environment with full functionality on web[2].

Picnic is a robust and commercial web tool. Unfortunately it is not intended as an IDE, However, there is the ACE project formerly known as SkyWritter and previously as Bespin, sponsored by the Mozilla Foundation, which aims to become one of the first IDE's online with support for different language. Despite being launched in 2008, it has faced several problems which have prevented its consolidation [14].

Finally we have the App Inventor, a tool based on Scratch and released in 2010 by Google. App Inventor is the best example of how an IDE web could be, since its implementation is in a browser and the projects carried out in it are kept on line. However, for it to carry out its performance you must download and install a module, without it, the inventor web app would not be functional.

One of the biggest advantages is that by being based on the Scratch project, it is similar in the way it implements the code, ie the implementation is given in blocks, all in a visually appealing, simple and functional way from a web browser [15].

---

[2] Picnik: http://www.picnik.com/

## 2. Future scenario.

### 2.1 New technologies

The new visual recognition technologies can help to build intuitive and simple gestural interaction innovative systems, although these technologies were designed for video game consoles, their use could be extended to personal computers, bringing a new trend of "natural " interaction [16].

This is because gestures with the hands such as: press a button, activate it with your finger, drag and drop, can be formalized in a kind of alphabet and used as a means of interaction [4].

With the release of the Kinect microsoft, development groups were formed to create free drivers. That is how the Libfreenect project arises, with the emergence of this driver, we had the opportunity to use kinect environments outside the video game console enabling new forms of interaction designed to complement the classic peripheral, such as the mouse and the keyboard[3].

### 2.2 Technologies involved.

To complete the project the combination of various technologies, is required. This is why the Libfreenect driver for Kinect is chosen as a viable basis for building the interaction system.

Libfreenect is an open source driver support for different operating systems which provide access to the RGB video stream kinect device.

By connecting the device to the USB interface its controller gets the video and audio streams, and conducts motor movements and changes in the receptor LED to each of the 5 possible states, thus generating a vector in memory. This vector is processed using the OpenCV library, this way, the handling of images is transparent to the developer.

As mentioned in 1.4 the Internet has not been used as a platform for building IDE's online. For such action communication between a browser and the device are required. Communication can be done through the library DepthJS, developed at the Media Lab at the Massachusetts Institute of Technology. The implementation is an extension for Google Chrome web browser, which allows communication with the kinect.

With the integration of these two technologies, the kinect would join the keyboard and mouse as an input device, creating a line of unconventional online IDE's gestures [4].

## 3. Description of the problem.

### 3.1 Definition of the problem.

Currently the development of code is not conceivable without the aid of tools to improve quality and reduce the time required to perform system development projects. Therefore, in the development phase the IDE plays a key role in the fulfillment of different tasks, ranging from code writing to its documentation.

These tools have evolved over time to become completely reliable, however trends have not been explored, using existing.

All of the IDE's development has been typecast towards the implementation and integration of more and better tools. Apparently the saturation of these tools in an IDE, goes unnoticed, since it is unlikely that any developer uses each and every one of them [3].

A possible improvement to solve the saturation is proposed by Eclipse with its philosophy of "extensibility", which plays an
important role because it allows customization through the integration of different plugins. Therefore, specific features are added, either to language or to carry out any other task as
documentation or diagrams for software design. In addition, Eclipse runs a sort of portability, although this support is based on the Java virtual machine, it only prevents installing the software on the local computer, because you can just download, unzip and use it.

However, this portability in practice is null, since it requires a distribution designed specifically for the platform on which to run.

### 3.2 Related work.

---

[3] Open Kinect: http://openkinect.org/wiki/Main_Page

As mentioned in the 1.3 The Discovery Method lacks of specialized tools to perform each of its phases. In 2009 Thom Parkers under the direction of Dr. Anthony J. H. Simmons developed the Discovery Method CASE Tool (DMCT), focused on the task design diagrams. Unfortunately the workflow editor is not yet completed, the functionality that users can use are as follows:

- Creation of diagrams.
- Export diagrams to JPG format.
- Printing of diagrams.

Despite not being a finished work it, can help understand the use of components of the diagram, it is notable that in 2009 some features were extended [17].

Another tool developed is the Task Flow Diagram Plug-in and Model checking Plug-in both implemented in Java, based on frameworks members of Eclipse IDE projects. It is for this reason that they have the characteristics mentioned in the 1.2 moreover, certain components of the Eclipse project, present difficulties to use it, creating delays and hindering the development [18].

## 4. Approximation to the solution.

### 4.1 Proposed Work.

We are proposing to explore a new form of interaction foe designing diagrams in an online IDE. For this purpose we are developing an integrated development environment, which has the ability to take as means of input the movements that the user performs with his hands, is to generate a new line of interaction. This is achieved by taking advantage of technologies that now allow the adding of a kind of natural communication by identification and establishment of certain patterns described by the movement of the hands.

Due to the nature of the project, we are using the user-centered design process in order to design and build a high-fidelity prototype.

Here we present the graph in Figure 1 showing the activities of the Unified Process with the UCD process fitted into the project life cycle.

In Figure 2 we can appreciate some of the activities of the UCD process such as:

- Interviews with users.
- Heuristic evaluation.
- Techniques such as card sorting.

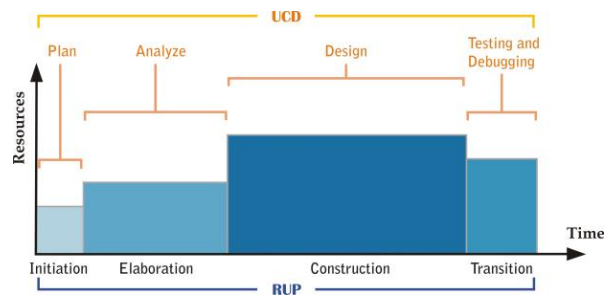

Figure 1. UCD phases into RUP [19].

It is because we are trying to offer an evolution of the user experience using a new technology that we think the UCD process and its coupling in the unified process [20, 19] could get a better user experience with minimal effort from the users [21].

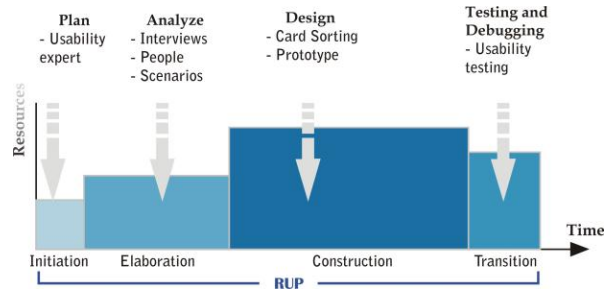

Figure 2. UCD activities coupled within the UP [19].

Our aproximation is targeting the use of a new technology: Kinect. Kinect shall be available as a new means of interaction complementary to mouse and keyboard. This will create an alphabet of motions described by patterns made with the movement of hands for manipulating diagrams.

To make this possible, part of the web environment design will be modified, integrating DepthJS library as a basis for communication with the kinect and its subsequent interpretation and comparison with the alphabet of movement.

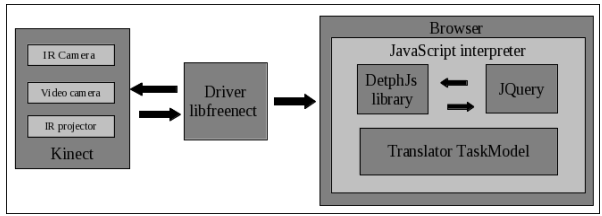

Figure 3. Architecture of the prototype.

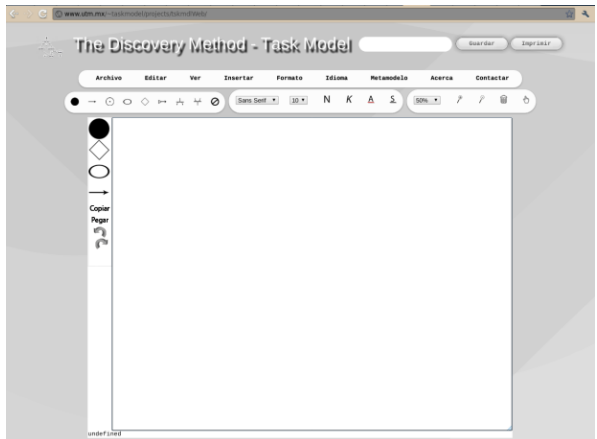

Figure 4. Low-fidelity web prototype.

The objectives for developing the project to have a high-fidelity prototype is described by the general purpose of manipulating diagrams through the free movement of hands in a web environment in order to integrate their own hands as a means of interaction.

Additionally the following points need to be fullfiled:

- Design a Web IDE proposal for manipulating diagrams.

- partial Development IDE proposed to obtain a high fidelity prototype.

- Creation by the free movements of the hands of an alphabet handling for diagrams.

- Measuring the user experience through usability testing.

### 4.2 Approximation to the solution.

The main objective of the research is "to manipulate charts using hands free movement" to do this various activities are considered to meet the phase-centered design phases of the user. The first activities to be carried out consist of a study of technologies to achieve the desired interaction, as well as to identify the system users. Following this stage, the aim is to obtain a low-fidelity prototype to implement the required parts and thus usability testing on a high-fidelity prototype.

Based on the results of usability testing, possible improvements to be made in both the design and the implementation, can be identified Figure 5 shows the phases making up the project.

## 5. Conclusions

Advances in technology now allow you to explore new forms of interaction, in fact, we currently have a large number of devices which lack traditional interaction technologies such as mechanical keyboards or cursor keys, and are instead controlled with natural elements by users, such as voice, touch and movement.

However, this modernity has not arrived to the developers themselves, because as in the early era of computing, the tools they use have not undergone significant changes with respect to devices and programs used by unskilled users in programming.

The advancement of this technology has enabled to have real mobility and portability. Few applications have based their operation on the site and thus closer to a higher level of platform independence.

Thus by studying the most natural mover and the capture and interpretation of these, we offer a truly new form of interaction, based on emerging technologies, which despite having little time on the market, enjoy a wide acceptance.

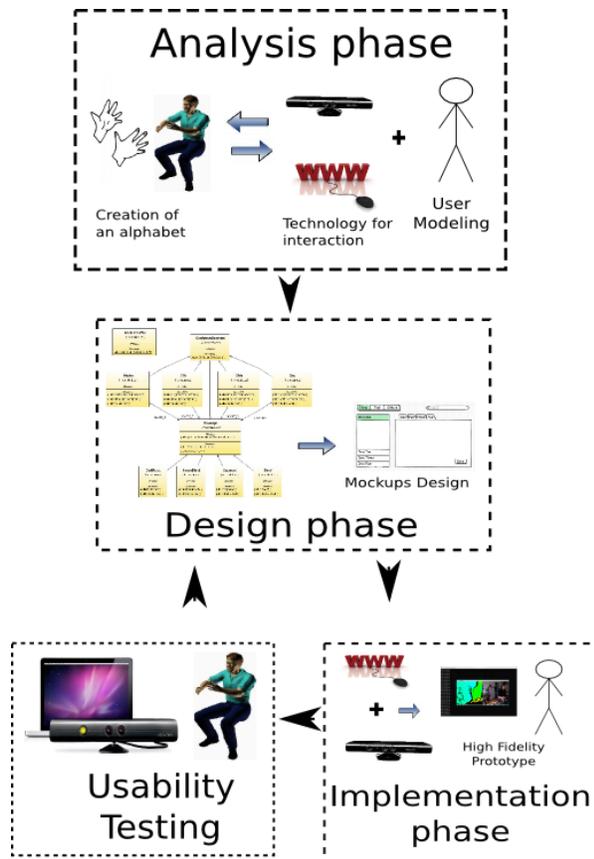

Figure 5: Descriptive diagram for the phases that make up the project development.

Acknowlegment

This work has been funded by the Universidad Tecnológica de la Mixteca.